\documentclass[a4paper,11pt]{article}
\usepackage{pos}

\title{Exclusive $J/\psi$ photoproduction off deuteron in d+Au ultra-peripheral collisions at STAR}
\author*[a]{Zhoudunming Tu, for the STAR Collaboration}
\affiliation[a]{Department of Physics, Brookhaven National Laboratory, Upton, NY 11973, USA}
\emailAdd{zhoudunming@bnl.gov}

\abstract{
Gluon density and its distributions inside nuclei and the parton modification of bounded
nucleons inside a nucleus, are some of the main standing problems in nuclear and particle physics. In recent years, ultra-peripheral collisions (UPC) of heavy ions have provided a new way of probing the gluon density, which is based on coherent diffractive vector-meson productions, e.g., $J/\psi$ meson. For heavy ions, e.g., Pb, the gluon density is found to be significantly suppressed through the UPC $J/\psi$ measurement, suggesting a strong gluon shadowing effect in heavy nuclei. In this analysis, we aim to look at a unique set of data taken by the STAR experiment, where $J/\psi$ mesons are photoproduced off the deuteron target with no other particle produced, except for the deuteron or its breakup products. The Zero Degree Calorimeter response with respect to the deuteron dissociation by detecting a beam-rapidity neutron is also investigated and provides additional information about the underlying physics process. The cross section of $J/\psi$ photoproduction in the photon-deuteron system is measured at the photon-nucleon center-of-mass energy $W\sim25~\rm{GeV}$, as well as the momentum transfer $t$ dependence cross section, $d\sigma/dt$. Data suggests a wider gluon density distribution than the Hulthen charge density distribution in deuteron. 
}

\FullConference{%
  HardProbes2020\\
  1-6 June 2020\\
  Austin, Texas}

\begin{document}
\maketitle
\section{Introduction}
One of the outstanding puzzles in nuclear physics is the EMC effect~\cite{Aubert:1983xm}. Initially discovered by the European Muon Collaboration, the bounded nucleons inside nucleus showed a significant suppression of their parton distributions in the valance quark regions. The measurement was done with respect to a deuterium target - assuming the deuteron is free of nuclear effect. How much the suppression is for the same kinematics depends on the Atomic mass (A) number, where heavier nucleus were found to be more suppressed than medium or light nucleus. In recent years, new experimental evidences suggest that the origin of EMC effect is deeply connected to the Short-Range Nucleon Correlations (SRC), i.e., significant off-shell nucleon pairs, which might exhibit huge nuclear suppression on parton density in medium and heavy nucleus~\cite{Higinbotham:2010tb}. However, alternative interpretations of the data are not excluded~\cite{Higinbotham:2010tb}, and this problem remains as a puzzle that is not definitively answered to-date. 

Besides the EMC effect in the region of valance quarks, the gluon density 
at low Bjorken-x region in bounded nucleons, has also been found to be 
different than free nucleons. In recent LHC experiments, photoproduction of 
vector-meson, e.g., $J/\psi$ meson, has been measured in ultra-peripheral collisions (UPC) of heavy ions, and its cross section is found to be significantly suppressed with respect to free nucleons~\cite{Khachatryan:2016qhq}. The modification of gluon density in heavy nucleus and its underlying physics mechanism are a subject of interest for a wide range of physics communities, going from nuclear and particle physics to studies of high density neutron stars in astrophysics~\cite{Ferreiro:2013pua,Annala:2019puf}. Hereby we propose to investigate the gluon density in the simplest two-body nuclear system  - deuteron, with coherent photoproduction of $J/\psi$ meson off deuteron in UPC. In addition, the momentum transfer distribution $-t$, which can be approximated by the transverse momentum squared ($p^{2}_{\rm T}$) of the $J/\psi$ particle in photoproduction, is measured in order to probe the gluon spatial distribution inside the deuteron. Similar measurement has been proposed~\cite{Tu:2020ymk} to perform the gluon imaging of bounded nucleons with and without novel short range correlations at the future Electron-Ion Collider. This analysis will be the first measurement of gluon imagining for the deuteron target. 
\section{Result}
\begin{figure}[thb]
\includegraphics[width=2.50in]{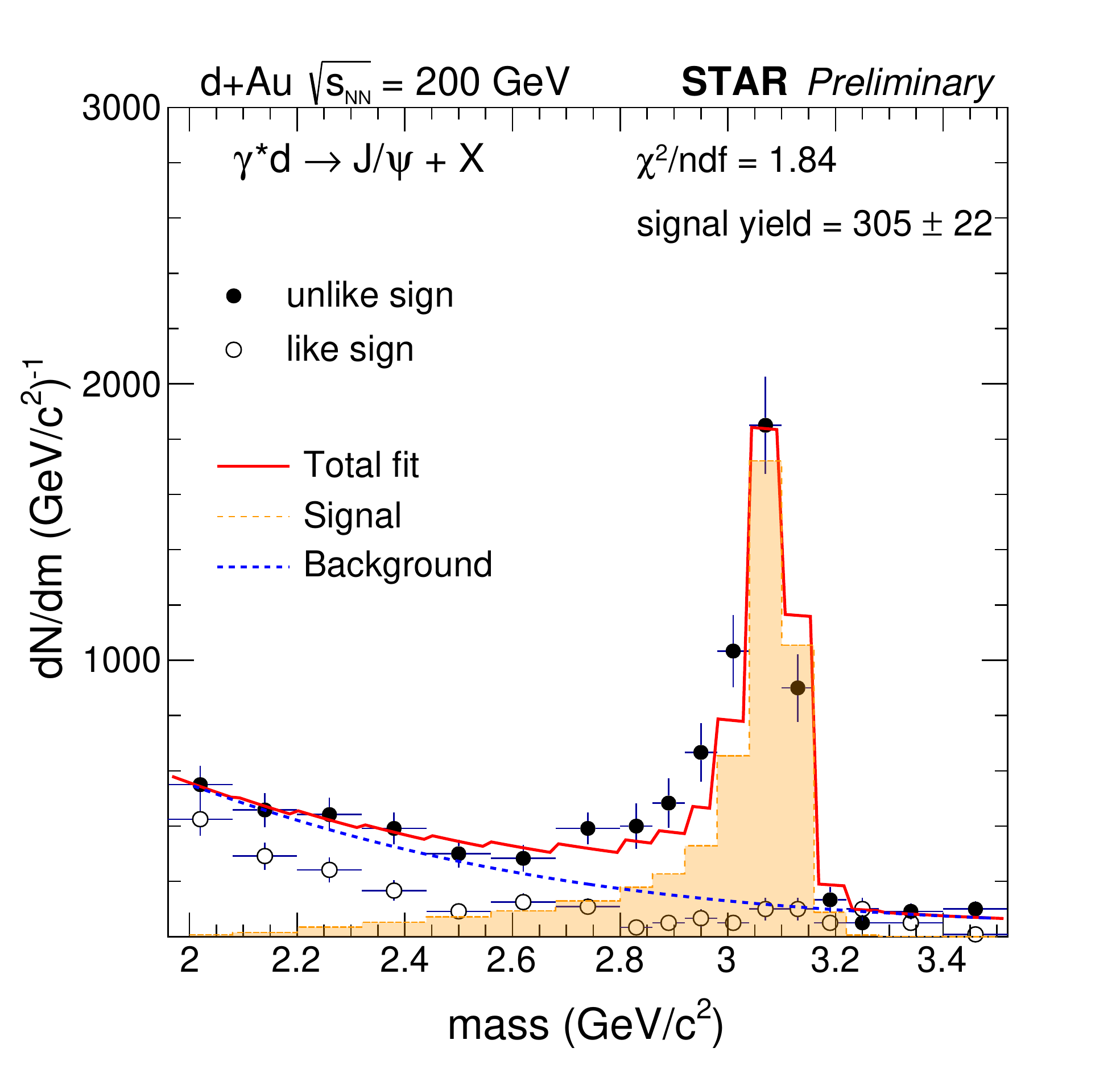}
\includegraphics[width=2.50in]{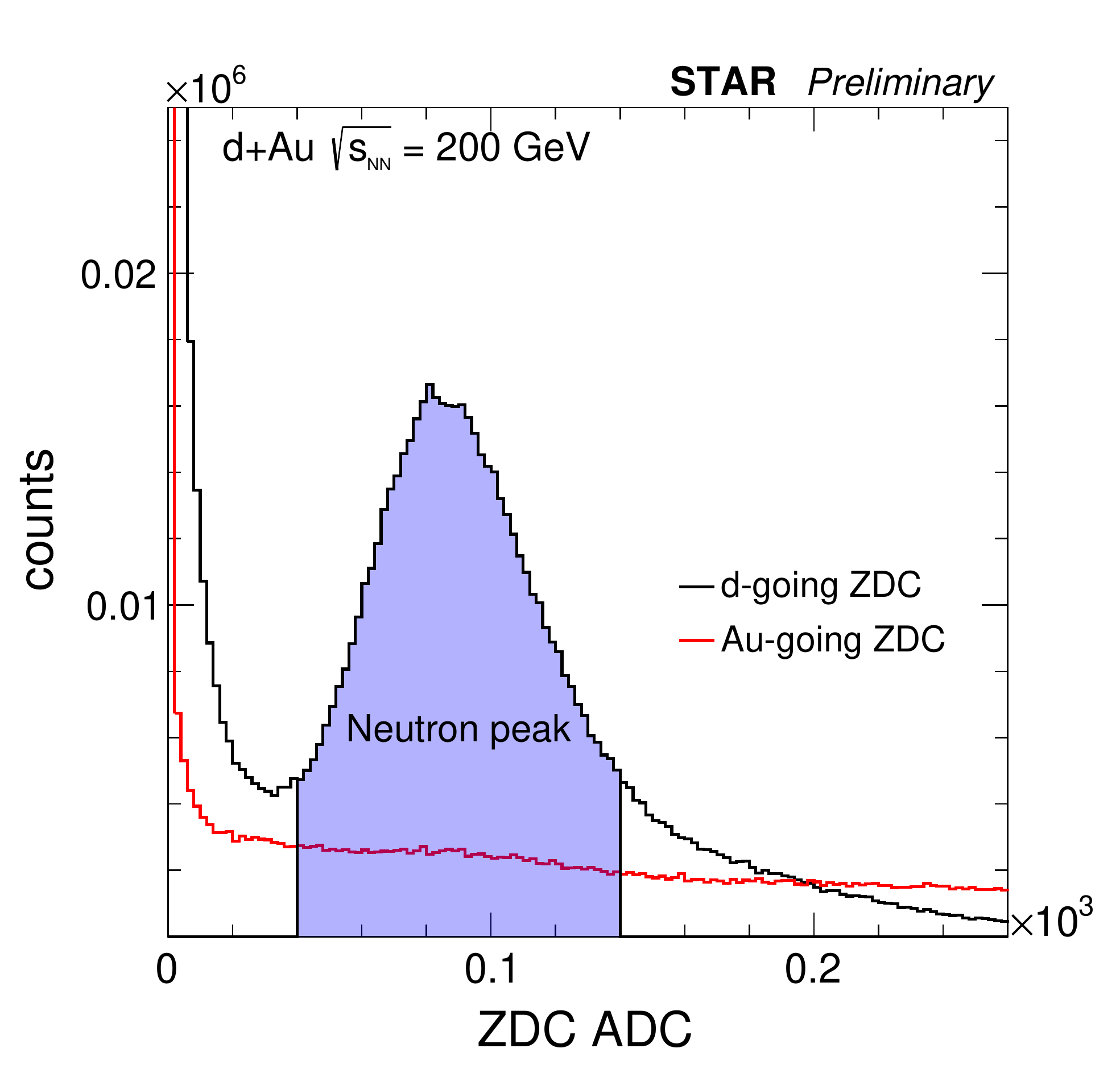}
  \caption{ \label{fig:figure_1}Left: Invariant mass peak distribution of d+Au UPC event around $J/\psi$ mass. Right: Zero Degree Calorimeter energy distribution for east and west side of STAR }
\end{figure}

\begin{figure}[thb]
\includegraphics[width=2.50in]{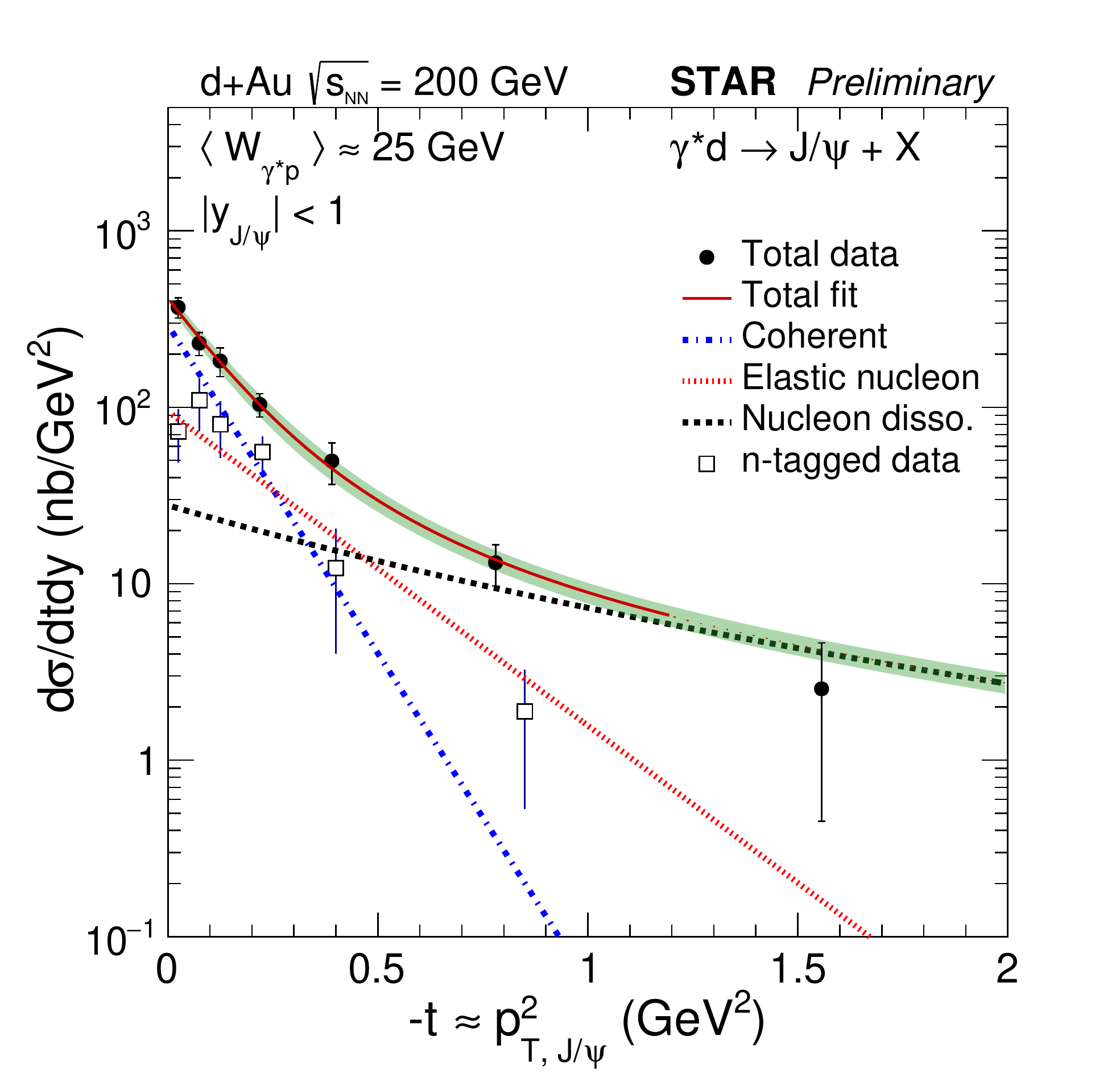}
\includegraphics[width=2.50in]{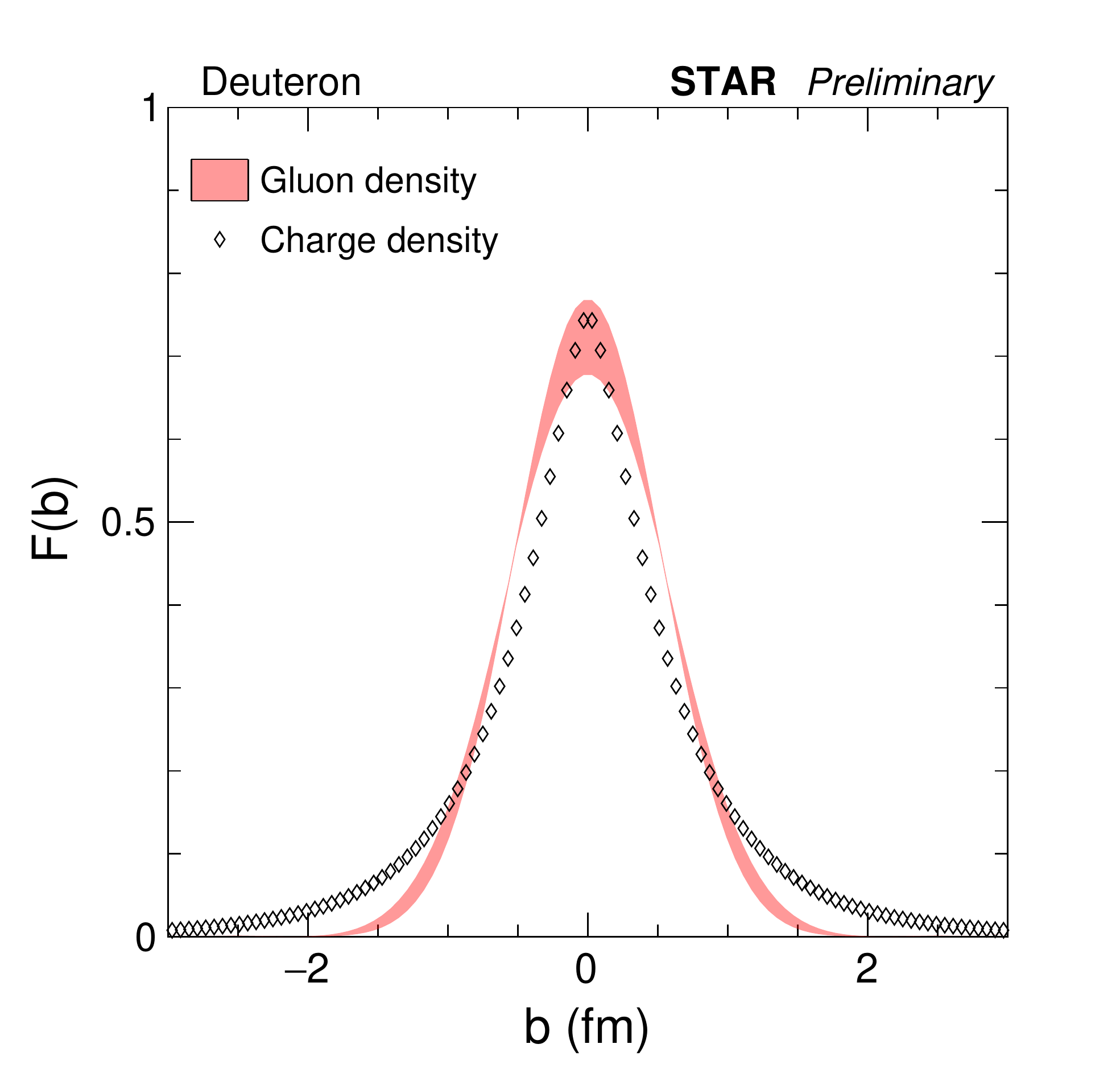}
  \caption{ \label{fig:figure_2}Left: Differential cross section of $J/\psi$ photoproduction as a function $p^{2}_{\rm{T}}$ is shown in $\rm d+Au$ UPC at STAR. Right: Fourier transformation from coherent $J/\psi$ photoproduction $t$ distribution to the spatial distribution in terms of impact parameter $b$ for the deuteron.}
\end{figure}

The online trigger for the enhanced UPC $J/\psi$ dataset was mainly based on the Barrel Electromagnetic Calorimeter response above roughly 0.5 GeV energy threshold for a back-to-back pair of towers and no other activity in the main detector.
In Fig.~\ref{fig:figure_1} left, the invariant mass distribution is shown for reconstructed the $J/\psi$ candidates. The total fit includes a signal template based on the GEANT simulation of $J/\psi$ signal and a background function to describe the data. The background function is $(x-A)e^{B(x-A)(x-C)+Cx^{3}}$, where $A$, $B$, and $C$ are fit parameters. The signal yield is extracted from the fit, which is used for calculation of the differential cross section. The definition of the differential cross section is the same as in Ref.~\cite{Acharya:2019vlb}. 
In Fig.~\ref{fig:figure_1} right, the Zero Degree Calorimeter (ZDC) ADC distribution is shown for both the east (Au-going) and the west (d-going) side of STAR detector. The Au-going side has no neutron peak, while the deuteron-going side is shown with a prominent neutron peak, indicating the deuteron has dissociated at least one neutron. All events shown are coincidence with a $J/\psi$ candidate, which is reconstructed via the di-electron channel at mid rapidity. 
In Fig.~\ref{fig:figure_2} left, the differential cross section of photoproduction $J/\psi$ off the deuteron as a function of $p^{2}_{T}$ of $J/\psi$ is shown for both with and without vetoing a neutron in the deuteron-going direction ZDC. The cross section is corrected by the transversely polarized photon flux such that the cross section is at the level of photon-deuteron system. The transversely polarized photon flux is calculated based on STARlight event generator~\cite{Klein:2016yzr}. The total differential cross section includes three components, the coherent diffractive production, incoherent diffractive but without breaking the nucleon, and the nucleon dissociations. The latter two components are almost equivalent to the elastic and nucleon dissociation of $J/\psi$ photoproduction off protons measured at HERA~\cite{Alexa:2013xxa}. 
In this analysis, the H1 data is used for a template fit, where the coherent diffractive component is assumed to be an exponential function and the slope is a fit parameter. The shape of incoherent and nucleon dissociation are fixed while leaving the absolute cross section a fit parameter. The extracted slope of the coherent diffractive component is $-8.5\pm1.2\rm{(stat.)}\pm1.5\rm{(sys.)}~\rm{GeV^{-2}}$, which is consistent with a naive expectation that the slope parameter is closely related to the deuteron size. Finally, by using this slope parameter, a Fourier transformation has been performed and the source distribution of gluon density $F(b)$ as a function of the impact parameter $b$ is shown in the Fig.~\ref{fig:figure_2} right. The data suggests a wider gluon density distribution than its charge density distribution. 
  
\section{Summary}
The photoproduction of $J/\psi$ vector meson off the deuteron has been measured for the first time by using the d+Au ultra-peripheral collisions at STAR. The differential cross section as a function of the $J/\psi$ transverse momentum squared is presented and the coherent diffractive component has been extracted based on template fit method. The incoherent and nucleon dissociation contributions are based on the H1 data of electron proton scattering. The gluon density distribution is found to be wider than the charge density distribution of the deuteron, which provides further insight into the gluon structure of light nuclei, e.g., deuteron, at high energy.
\bibliographystyle{aps}
\bibliography{reference}% Produces the bibliography via BibTeX.

\end{document}